\documentclass[preprintnumbers, floatfix,letterpaper,aps,prd,epsfig,nofootinbib,
twocolumn
]{revtex4-1}
\pdfoutput=1
\usepackage{bm,graphicx,dcolumn,epstopdf,epsf, latexsym,mathbbol, amssymb,amsmath,color,slashed, mathrsfs,mathcomp, simplewick}
\usepackage{multirow}
\usepackage{makecell}
\usepackage{lmodern}
\usepackage[T1]{fontenc}
\usepackage[latin9]{inputenc}
\usepackage{subfig}
\usepackage{xcolor}
\usepackage{babel}
\usepackage{textcomp}
\usepackage{amsmath}
\usepackage{amsthm}
\usepackage{amssymb}
\usepackage{graphicx}
\PassOptionsToPackage{normalem}{ulem}
\usepackage{ulem}
\usepackage[colorlinks,linkcolor=blue,citecolor=blue,urlcolor=blue]{hyperref}

\begin{document}
	\allowdisplaybreaks
	\newcommand{\bq}{\begin{equation}}
	\newcommand{\eq}{\end{equation}}
	\newcommand{\bqn}{\begin{eqnarray}}
	\newcommand{\eqn}{\end{eqnarray}}
	\newcommand{\nb}{\nonumber}
	\newcommand{\lb}{\label}
	\newcommand{\f}{\frac}
	\newcommand{\p}{\partial}
	\newcommand{\PRL}{Phys. Rev. Lett.}
	\newcommand{\PLB}{Phys. Lett. B}
	\newcommand{\PRD}{Phys. Rev. D}
	\newcommand{\CQG}{Class. Quantum Grav.}
	\newcommand{\JCAP}{J. Cosmol. Astropart. Phys.}
	\newcommand{\JHEP}{J. High. Energy. Phys.}
	\newcommand{\red}{\textcolor{red}}
	
\title{The effects of asymptotically flat ${\cal R}^{2}$ spacetime on black hole image of Sagittarius A* }

\author{Jian-Ming Yan${}^{a, b}$}
\email{yanjm@zjut.edu.cn}

\author{Qiang Wu${}^{a, b}$}
	\email{wuq@zjut.edu.cn}
   
\author{Tao Zhu${}^{a, b}$}
	\email{zhut05@zjut.edu.cn; Corresponding author}

\affiliation{${}^{a}$Institute for Theoretical Physics \& Cosmology, Zhejiang University of Technology, Hangzhou, 310023, China\\
${}^{b}$ United Center for Gravitational Wave Physics (UCGWP),  Zhejiang University of Technology, Hangzhou, 310023, China\\}
	
\date{\today}
	
\begin{abstract}

A new class of analytically expressible vacuum solutions has recently been discovered for pure ${\cal R}^2$ gravity, building upon Buchdahl's seminal work from 1962. These solutions, inspired by Buchdahl's framework, offer a promising avenue for testing ${\cal R}^2$ gravity against astrophysical observations. Within a subset of asymptotically flat Buchdahl-inspired vacuum spacetimes, we introduce a free parameter $\epsilon$ to characterize deviations from the Schwarzschild metric, which is recovered in the limit $\epsilon = 0$. In this study, we employ the publicly available code \textit{ipole} to simulate black hole images under the Buchdahl-inspired metric, with a focus on the black hole at the center of the Milky Way, Sagittarius A* (Sgr A*). Our simulations show that both the shadow size and photon ring diameter decrease monotonically with increasing $\epsilon$. By exploring a range of observational inclination angles, we find that the photon ring diameter being a direct observable is only weakly sensitive to the inclination angle. We further constrain the parameter $\epsilon$ by comparing our simulation results with the Event Horizon Telescope (EHT) observations of Sgr A*. The obtained bounds are consistent with those previously derived from the orbital motion of the S2 star, but provide tighter constraints. In addition, we analyze the influence of the Buchdahl-inspired spacetime on the polarization patterns near the black hole and find its impact to be minimal. In contrast, the observational inclination angle has a substantial effect on the observed polarization structure, highlighting the dominant role of viewing geometry in shaping polarization features.
\end{abstract}

\maketitle

\section{Introduction}

Over the past century, Einstein's General Relativity (GR) has established itself as an extraordinarily successful theory of gravitation, providing precise explanations for the behavior of massive objects under gravitational influence. It has consistently delivered accurate predictions for various observed phenomena, such as the precession of Mercury's perihelion, gravitational redshift effects, and the Shapiro time delay \cite{Will:2014kxa, Park:2017zgd, Fomalont:2009zg, Cassini, Stairs:2003eg}. More notably, GR has predicted the existence of remarkable gravitational phenomena, including black holes, compact stars, and gravitational waves, well before they were directly observed. In recent decades, advanced observational campaigns have resulted in high-resolution images of black hole shadows, such as those at the centers of M87 and the Milky Way \cite{EHT1, EHT2, EHT3, EventHorizonTelescope:2021bee, EHT4,EventHorizonTelescope:2022wkp}. Additionally, numerous gravitational wave signals from binary black hole mergers with varying masses have been successfully detected \cite{gws, gws2}, further confirming the validity of GR. As a result, GR remains unmatched as the only gravitational theory to have passed all tests across both solar system and astrophysical scales.

Despite its outstanding successes, GR faces significant challenges. Key among these are the need for renormalization and the quest to reconcile GR with quantum mechanics, which aims to develop a unified framework capable of accurately describing phenomena across both microscopic and cosmological scales.

These challenges suggest the potential necessity of modifying GR. In the pursuit of a quantum gravity theory, string theory has emerged as a promising candidate, offering a unified description of gravity and quantum mechanics. The effective action of string theory not only incorporates the Einstein-Hilbert action but also includes higher-order curvature corrections. Studies of the ${\cal R}+{\cal R}^2$ gravity model have demonstrated that ${\cal R}^2$ gravity possesses desirable properties, such as scale invariance, absence of ghost instabilities, renormalizability, and the potential for embedding within supergravity \cite{Gaume-2016}. Consequently, it has emerged as a viable candidate for modified gravity and quantum gravity theories. However, analytically solving the gravitational equations for ${\cal R}^2$ gravity remains a significant challenge, as the field equations involve fourth-order derivatives of the metric tensor, making exact solutions elusive.

In earlier work, Buchdahl attempted to find new static, spherically symmetric solutions to the spacetime or black hole equations within the framework of the ${\cal R}^2$ theory~\cite{Buchdahl-1962}. His efforts led to the derivation of a second-order ordinary differential equation that addressed longstanding challenges related to the metric coefficients. More recently, Nguyen made significant progress in this direction by successfully obtaining vacuum solutions within the same framework~\cite{Nguyen-2022-Buchdahl, Nguyen-2022-Lambda0, Nguyen-2023-essay, 2023-WH, 2023-axisym, taohoang, Jusufi:2020cpn, Liu:2020ola, Zhu:2019ura,Shi:2024bpm}. For asymptotically flat spacetimes, the resulting metric introduces a free parameter $\epsilon$, which encapsulates the influence of higher-derivative corrections. When $\epsilon = 0$, the solution naturally reduces to the Schwarzschild metric. 

In~\cite{taohoang}, the authors tested the Buchdahl-inspired spacetime against several classical experiments, finding that the parameter $\epsilon$ is tightly constrained to values near zero. In our previous work~\cite{Yan:2024bim}, we further constrained this parameter using the orbital dynamics of the S2 star, obtaining a range of $\epsilon \in (-0.6690, 0.4452)$. Unlike tests performed within the solar system, the environment surrounding the S2 star is characterized by a much stronger gravitational field, providing a valuable regime for probing potential deviations from GR.

In 2022, the Event Horizon Telescope (EHT) collaboration released the first image of the shadow of the supermassive black hole at the center of our galaxy,  Sagittariu(Sgr A*)~\cite{EventHorizonTelescope:2022wkp}. This milestone, following the earlier image of M87*, marks the second direct image of a black hole and opens a new window into the study of black holes and their extreme environments~\cite{Psaltis:2018xkc, Vagnozzi:2022moj, Perlick:2021aok, Allahyari:2019jqz, Gralla:2020srx, Cardoso:2018ptl, Volkel:2020xlc, Yan:2019hxx, Chen:2021lvo, Chen:2022nbb, Afrin:2022ztr, Kuang:2022ojj, Hou:2022eev, Jiang:2023img}. The shadow of Sgr A* carries significant scientific implications---not only does it offer further confirmation of GR, but it also provides novel opportunities to probe black hole accretion processes, jet formation mechanisms, and gravitational lensing effects.

Motivated by these developments, we aim to explore the observational consequences of the Buchdahl-inspired spacetime by conducting numerical simulations of black hole images, using Sgr A* as a representative target. By systematically varying the deformation parameter $\epsilon$, we investigate how this modified gravity framework influences observable features, with particular attention to the photon ring diameter and polarization patterns. Furthermore, we account for the impact of the observer's inclination angle by simulating black hole images under different observational geometries.

Throughout this paper, we adopt units where $c = 1$.

\section{A BRIEF INTRODUCTION OF THE GENERALIZED BUCHDAHL-INSPIRED METRICS}
\renewcommand{\theequation}{2.\arabic{equation}} \setcounter{equation}{0}

In this section, we present the generalized Buchdahl-inspired spacetime that is asymptotically flat, expressed in standard coordinates, and examine the motion of massive test particles within this background. Using the coordinate system $(t, R, \theta, \phi)$, the spacetime metric takes the Morris-Thorne form \cite{MorrisThorne}:
\begin{equation}
ds^2 = -e^{2\Phi(R)} dt^2 + \frac{dR^2}{1 - \frac{b(R)}{R}} + R^2 d\Omega^2. \label{R_coors}
\end{equation}
The functions $\Phi(R)$ and $b(R)$, which govern the redshift and shape of the geometry respectively, are detailed in \cite{2023-WH}. The areal radius $R$ is defined in terms of an auxiliary variable $y$ through the following relation:
\begin{equation}
R = \zeta r_s \frac{y^{\frac{\tilde{k} - 1}{\zeta} + 1}}{1 - y^2}. \label{eq:areal-radius}
\end{equation}
Correspondingly, the redshift and shape functions are expressed as:
\begin{align}
e^{2\Phi(R)} &= y^{\frac{2}{\zeta}(\eta + 1)}, \label{eq:redshift-func} \\
1 - \frac{b(R)}{R} &= \frac{1}{4y} \left[ (y^2 + 1) + \frac{\tilde{k} - 1}{\zeta}(1 - y^2) \right]^2. \label{eq:shape-func}
\end{align}
When the parameters are chosen as $\eta = \tilde{k}$ and $\zeta = \sqrt{1 + 3\tilde{k}^2}$, the resulting metric \eqref{R_coors}-\eqref{eq:shape-func} coincides with the asymptotically flat solution originally derived in \cite{Nguyen-2022-Lambda0}. Moreover, the particular case $\zeta = 1$ recovers the Campanelli-Lousto solution introduced in \cite{CampanelliLousto-1993}.

In the weak-field regime, where gravitational fields are relatively small, the metric \eqref{R_coors} simplifies to:
\begin{align}
ds^2 \approx & -\left(1 - \frac{(1 + \eta) r_s}{R} + \frac{(1 + \eta)(\tilde{k} + \eta) r_s^2}{2 R^2}\right) dt^2 \nonumber \\
& + \left(1 + \frac{(1 - \tilde{k}) r_s}{R}\right) dR^2 + R^2 d\Omega^2. \label{weak}
\end{align}
Here, $\eta$ and $\tilde{k}$ are treated as free parameters. The Schwarzschild radius $r_s$ relates to the black hole mass $M$ via $r_s = 2GM$, where $G$ denotes the gravitational constant. A comparison with the Newtonian limit leads to a redefinition of $G$ in terms of the Newtonian gravitational constant $G_{\rm N}$, given by $(1 + \eta)G = G_{\rm N}$. Consequently, the $g_{tt}$ component becomes:
\begin{equation}
g_{tt} = -\left(1 - \frac{2 G_{\rm N} M}{R} + 2 \frac{\tilde{k} + \eta}{1 + \eta} \frac{G_{\rm N}^2 M^2}{R^2} \right).
\end{equation}

To facilitate the analysis, we introduce a dimensionless parameter $\epsilon$, which quantifies deviations from the Schwarzschild solution:
\begin{equation}
\epsilon \equiv \frac{\tilde{k} + \eta}{1 + \eta}, \quad G_{\rm N} = 1, \quad R \to r.
\end{equation}
In terms of this new parameter, the weak-field metric simplifies to:
\begin{eqnarray}
    ds^2 &\approx &-\left(1 - \frac{2M}{r} + \epsilon \frac{2M^2}{r^2} \right) dt^2 \nonumber\\
    &&+ \left(1 + \frac{2(1 - \epsilon) M}{r} \right) dr^2 + r^2 d\Omega^2. \label{newton}
\end{eqnarray}
This form represents the weak-field approximation of the Buchdahl-inspired spacetime. In the special case of $\epsilon = 0$, one recovers the canonical Schwarzschild metric of general relativity.

The location of the event horizon is determined by the condition that $g_{tt} = 0$. From equation \eqref{newton}, this yields:
\begin{equation}
g_{tt} = -\left(1 - \frac{2M}{r} + \epsilon \frac{2M^2}{r^2} \right).
\end{equation}
Solving this equation for the radial coordinate $r$ gives the horizon radius $r_{\text{h}}$:
\begin{equation}
r_{\text{h}} = M + \sqrt{M^2 - 2M^2 \epsilon}.
\end{equation}
As expected, when $\epsilon \rightarrow 0$, this expression reproduces the Schwarzschild radius $r = 2M$. For nonzero $\epsilon$, the event horizon is shifted, indicating a departure from standard GR and the influence of higher-curvature corrections inherent in the modified theory.

\section{The Accretion Flow}
\renewcommand{\theequation}{3.\arabic{equation}} \setcounter{equation}{0}

Given the low accretion rate and luminosity of the Sgr A* black hole, we adopt the radiatively inefficient accretion flow (RIAF) model\cite{Pu:2016qak}, which is characterized by its collisionless and low-density nature, as a more suitable accretion model for this scenario. In this study, we generate black hole images based on a semi-analytical RIAF model within a Buchdahl-inspired spacetime. The motion of the accretion flow is governed solely by the gravitational field of the black hole, allowing us to investigate its dynamics by solving the timelike geodesic equations in the Buchdahl-inspired black hole spacetime.

We begin by considering the Lagrangian for a massive particle:
\bq
\mathcal{L} = \frac{1}{2} g_{\mu\nu} \frac{dx^\mu}{d\lambda} \frac{dx^\nu}{d\lambda},
\eq
where $\lambda$ is the affine parameter. For timelike geodesics, the affine parameter $\lambda$ corresponds to the proper time $\tau$. From this Lagrangian, we derive the generalized momenta:
\bqn
p_{t} &=& g_{tt} \dot{t} = -E, \\
p_{\phi} &=& g_{\phi \phi} \dot{\phi} = L, \\
p_{r} &=& g_{rr} \dot{r}, \\
p_{\theta} &=& g_{\theta \theta} \dot{\theta},
\eqn
where $E$ represents the energy, $L$ is the angular momentum, and a dot denotes differentiation with respect to proper time $\tau$. Thus, we obtain the following relations:
\bq
\dot{t} = -\frac{E}{g_{tt}}, \quad \dot{\phi} = \frac{L}{g_{\phi \phi}}.
\eq

For simplicity, we focus on the motion of particles in the equatorial plane, where $\theta = \frac{\pi}{2}$ and $\dot{\theta} = 0$. For timelike particles, the condition $g_{\mu\nu} \dot{x^\mu} \dot{x^\nu} = -1$ holds. From this, we derive the radial equation of motion:
\bq
g_{rr} \dot{r}^2 - V_{\rm eff}(r) = 0,
\eq
where the effective potential $V_{\rm eff}(r)$ is given by:
\bq
V_{\rm eff}(r) = -\frac{g_{\phi\phi} E^2 + g_{tt} L^2}{g_{tt} g_{\phi\phi}} - 1.
\eq

\begin{figure}
\centering
\includegraphics[width=0.5\textwidth]{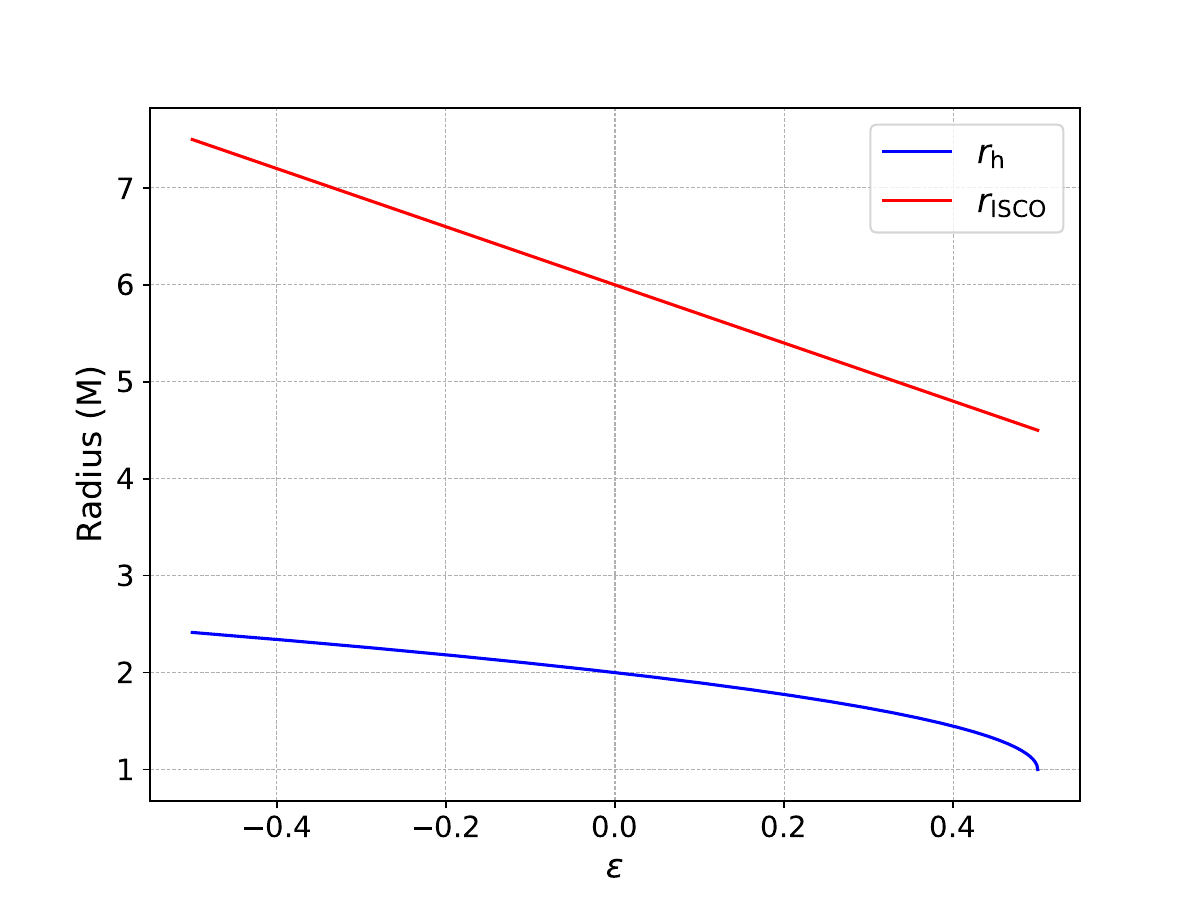}
\caption{This picture shows the trends of the event horizon radius $r_{\rm h}$ and the innermost stable circular orbit  radius $r_{\rm ISCO}$ as the parameter $\epsilon$ varies. The  blue line represents the  $r_{\rm h}$ and the red line represents the $r_{\rm ISCO}$. It is clear that both decrease as $\epsilon$ increases. } \label{isco}
\end{figure}

Next, we consider the conditions $\dot{r} = 0$, $\frac{dV_{\rm eff}}{dr} = 0$, and $\frac{d^2 V_{\rm eff}}{dr^2} = 0$, which allow us to determine the radius of the innermost stable circular orbit (ISCO) as:
\bq
r_{\rm ISCO} \simeq 6M \left(1 - \frac{1}{2} \epsilon\right).
\eq
As shown in FIG. \ref{isco}, both the event horizon and the ISCO decrease as $\epsilon$ increases.

For material outside the ISCO, the Keplerian angular velocity $\Omega_{\rm K}$ of the accretion disk is given by:
\bq
\Omega_{\rm K} = \frac{d\phi}{dt} = - \sqrt{\frac{-g_{tt, r}}{g_{\phi\phi, r}}}.
\eq

However, once a particle crosses the ISCO, its motion becomes increasingly complex. Stable orbits no longer exist, and the particle spirals inward, eventually being absorbed by the black hole.

To investigate the properties of sub-Keplerian orbits in the RIAF model, we refer to the approach in \cite{Pu:2016qak} and treat the flow as a dynamical model composed of both free-fall motion and Keplerian rotation.

First, we define the sub-Keplerian radial velocity $u^{r}_{\rm sub-K}$ as a combination of the Keplerian value $u^{r}_{\rm K}$ and the free-fall value $u^{r}_{\rm ff}$:
\bq
u^{r}_{\rm sub-K} = u^{r}_{\rm K} + (1-\alpha)(u^{r}_{\rm ff} - u^{r}_{\rm K}),
\eq
where $\alpha$ controls the deviation from Keplerian motion. Similarly, the sub-Keplerian angular frequency $\Omega_{\rm sub-K}$ is defined as:
\bq
\Omega_{\rm sub-K} = \Omega_{\rm K} + (1 - \beta)(\Omega_{\rm ff} - \Omega_{\rm K}),
\eq
where $\beta$ controls the deviation from purely Keplerian behavior. The Keplerian motion corresponds to $\alpha = \beta = 1$, while free-fall motion corresponds to $\alpha = \beta = 0$. In this study, we set $\alpha = \beta = 0.5$ to describe the sub-Keplerian motion.

To characterize the geometrical thickness of the accretion flow, we define a dimensionless scale height $H \equiv h / R$, where $h$ is the vertical scale height and $R$ is the radial coordinate. For our accretion flow model, which adopts a MAD-like (Magnetically Arrested Disk) configuration, we follow the convention in~\cite{Chen:2021lvo} and set $H = 0.3$. The normalized electron number density is set to $n_e = 3 \times 10^7 \, \text{cm}^{-3}$, and the electron temperature is fixed at $T_e = 3 \times 10^{11} \, \text{K}$. These values are motivated by modeling the supermassive black hole Sgr A* at the center of the Milky Way, and are chosen to reproduce the observed flux at 230~GHz in the range of 1--3~Jy~\cite{Jiang:2023img}. The black hole mass is taken to be $M_\bullet = 4 \times 10^6 \, M_{\odot}$, consistent with current observations, and the distance to the source is assumed to be 8~kpc.

\section{SIMULATION OF BLACK HOLE IMAGES UNDER THE GENERALIZED BUCHDAHL-INSPIRED METRICS}
\renewcommand{\theequation}{4.\arabic{equation}} \setcounter{equation}{0}

In this section, we shall discuss the black hole images we simulated under the Buchdahl-inspired spacetime. To compare with the observation of black hole by EHT, we use the Sgr A* as a target to do simulations. We use a public code \textit{ipole}\cite{Moscibrodzka:2017lcu} to generate images under different values of the parameter $\epsilon \in (-0.4,0.4)$ to study the effects of the Buchdahl-inspired spacetime on the black hole images of Sgr A*.

\subsection{Effects on the Photon Ring Diameter}

In this subsection, we investigate in detail how the parameter $\epsilon$ influences the photon ring diameter in the context of the Buchdahl-inspired spacetime. The photon ring diameter is a directly observable quantity, defined as the distance between the two prominent peaks in the flux distribution along a horizontal cross-section of the image. This geometric feature serves as a robust and model-independent observable that can be used to test deviations from GR.

\begin{figure*}
 \includegraphics[width=0.32\textwidth]{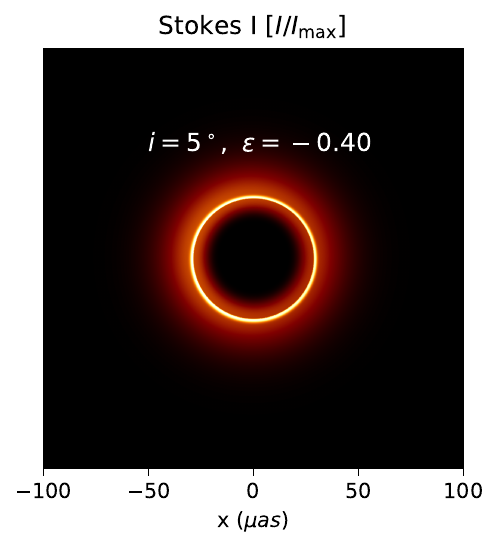}
 \includegraphics[width=0.32\textwidth]{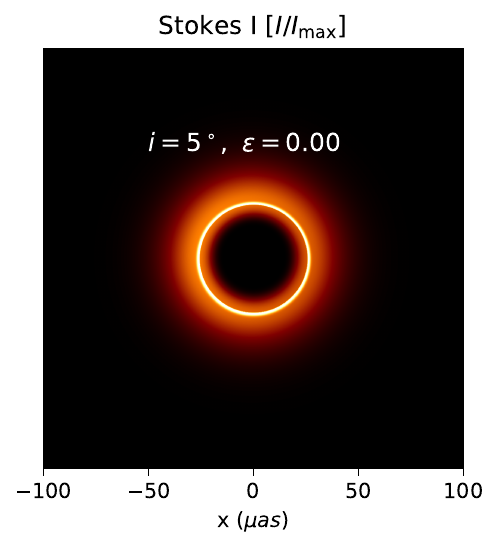}
 \includegraphics[width=0.32\textwidth]{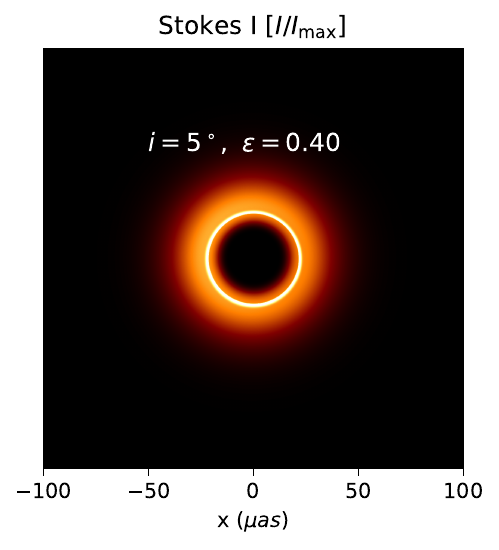}
 \includegraphics[width=0.32\textwidth]{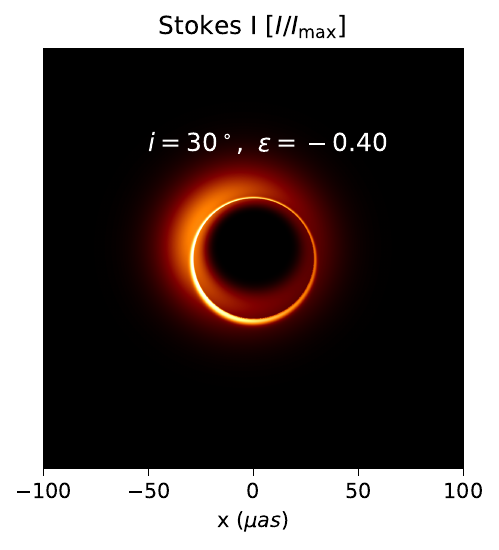} 
 \includegraphics[width=0.32\textwidth]{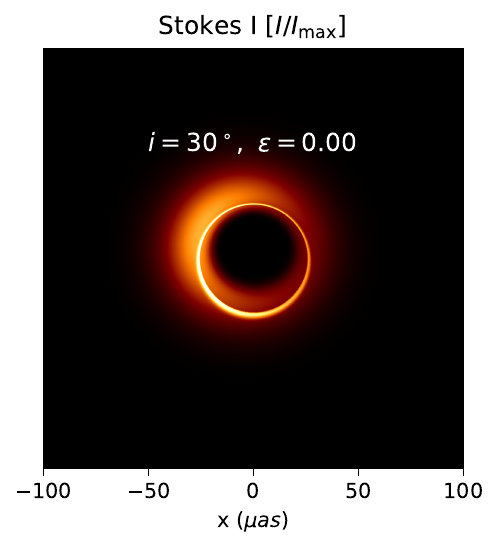}
 \includegraphics[width=0.32\textwidth]{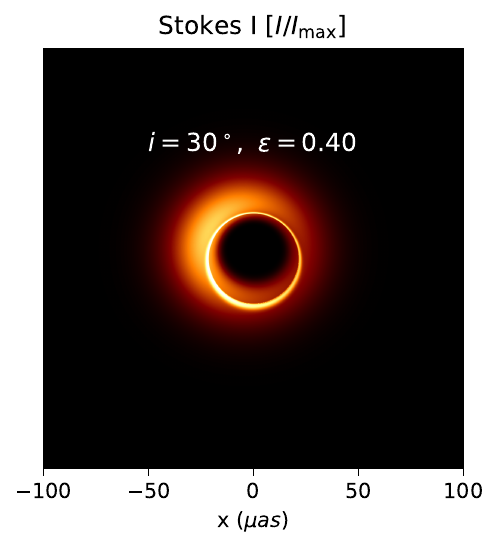}
 \includegraphics[width=0.32\textwidth]{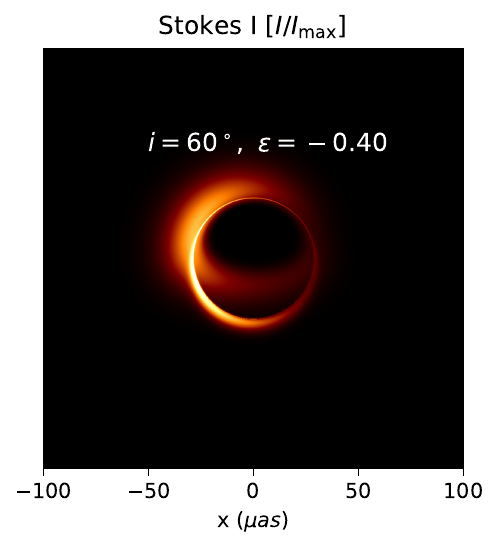}
 \includegraphics[width=0.32\textwidth]{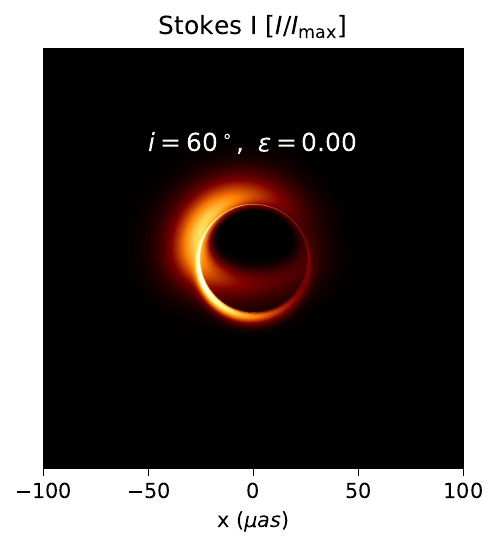}
 \includegraphics[width=0.32\textwidth]{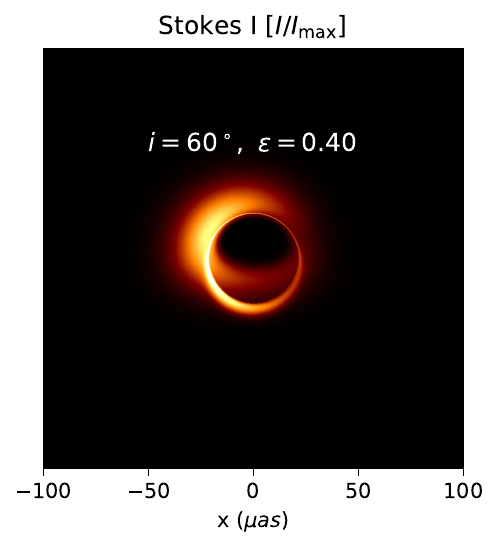}
\caption{The black hole shadow images for three different values of $\epsilon = -0.4$, $0$, and $0.4$ are shown for observation inclination angles of $5^\circ$, $30^\circ$ and $60^\circ$. The top row displays the images at an inclination of $5^\circ$, the middle row corresponds to an inclination of $30^\circ$, and the bottom row corresponds to an inclination of $60^\circ$. We observe that as the inclination angle increases, the brightness distribution in the black hole images exhibits more pronounced shifts and asymmetries. Nevertheless, the size of the black hole shadow and the photon ring diameter consistently decrease with increasing $\epsilon$, in agreement with the previous results.
}
    \label{bh}
\end{figure*}

\begin{figure*}
    \includegraphics[width=0.3\textwidth]{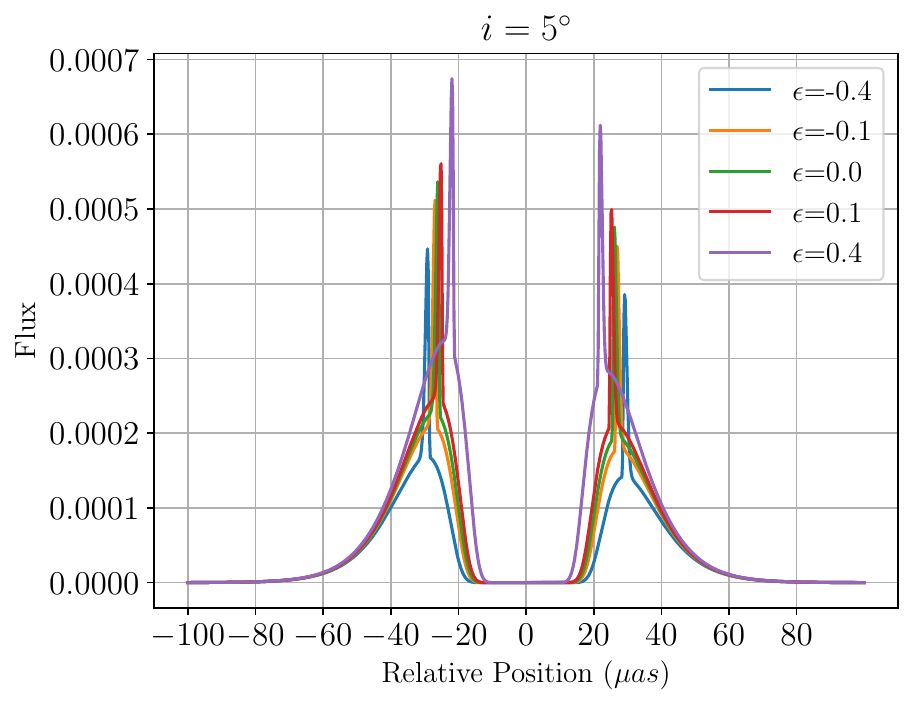}
    \includegraphics[width=0.3\textwidth]{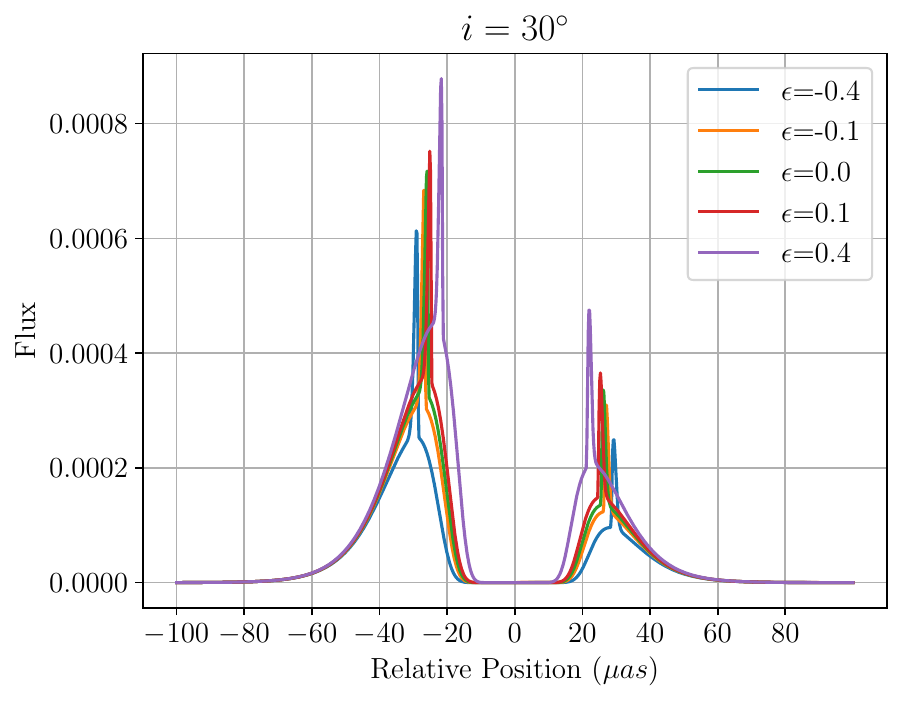}
    \includegraphics[width=0.3\textwidth]{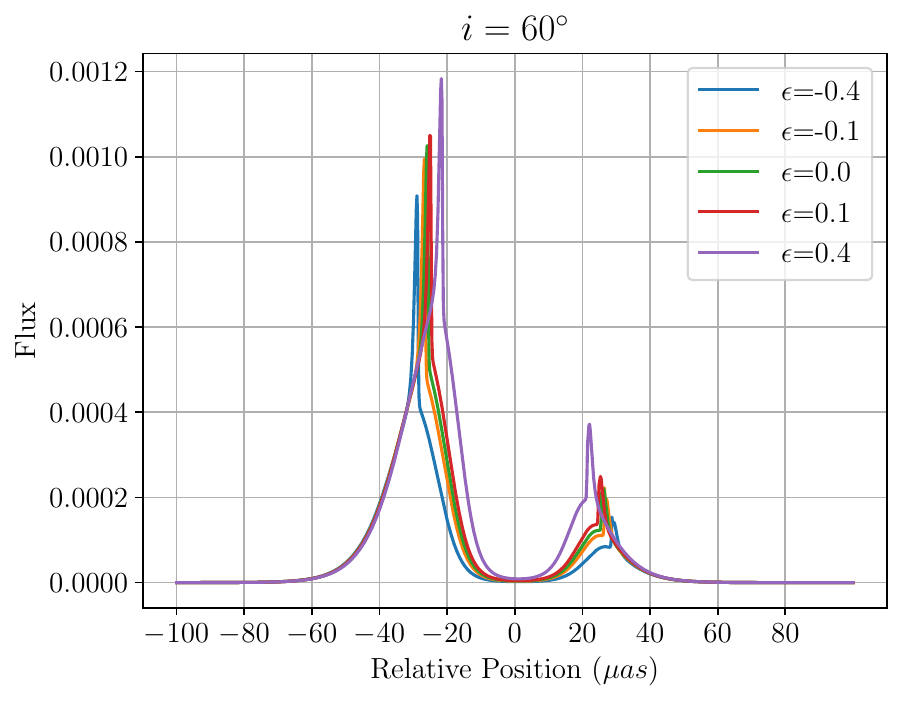}
 \caption{The flux distribution on the cross-cut profile with different value of $\epsilon$ under an observation inclination of $5^\circ$, $30^\circ$ and $60^\circ$. It is clear that the distance between the two peaks of flux decrease as $\epsilon$ increase. And the left-right symmetry of the flux distribution changes with varying observation inclination.}
    \label{flux}
\end{figure*}

In the previous section, we analyzed the dependence of the horizon radius $r_{\rm h}$ and the ISCO radius $r_{\rm ISCO}$ on the deformation parameter $\epsilon$, and found that both $r_{\rm h}$ and $r_{\rm ISCO}$ systematically decrease as $\epsilon$ increases. Given that the photon ring typically lies near or slightly outside the ISCO, it is reasonable to expect that the photon ring diameter would also exhibit a decreasing trend with increasing $\epsilon$.

Figure~\ref{bh} presents simulated black hole images in the Buchdahl-inspired spacetime for representative values of $\epsilon = -0.4$, $0$, and $0.4$, observed at inclination angles of $5^\circ$, $30^\circ$, and $60^\circ$, respectively. As shown in these images, both the size of the black hole shadow and the angular diameter of the photon ring decrease with increasing $\epsilon$, in agreement with theoretical expectations. Moreover, we observe that as the inclination angle increases, the alignment between the photon ring and the surrounding emission halo becomes more asymmetric. This increasing asymmetry in the brightness distribution highlights the role of relativistic beaming and gravitational lensing effects at high inclinations.

To further quantify the relationship between the photon ring diameter, the deformation parameter $\epsilon$, and the inclination angle, we analyze the one-dimensional flux profiles along the horizontal cross section for several representative values: $\epsilon = -0.4$, $-0.1$, $0$, $0.1$, and $0.4$. As shown in Fig.~\ref{flux}, the separation between the flux peaks corresponding to the photon ring diameter decreases monotonically with increasing $\epsilon$ for all three inclination angles. In addition, the left-right symmetry of the flux distribution becomes increasingly distorted at higher inclinations, illustrating the significant impact of viewing geometry on the image morphology.

\begin{figure*}
    \includegraphics[width=0.3\textwidth]{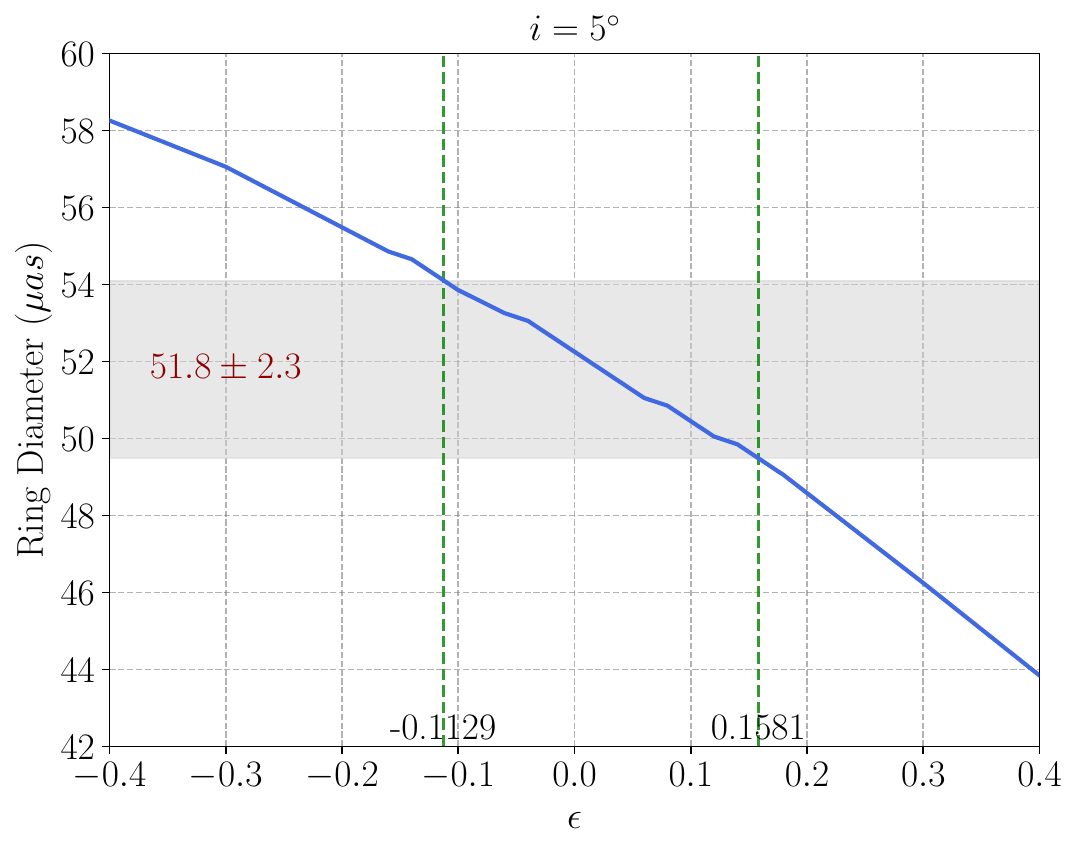}
    \includegraphics[width=0.3\textwidth]{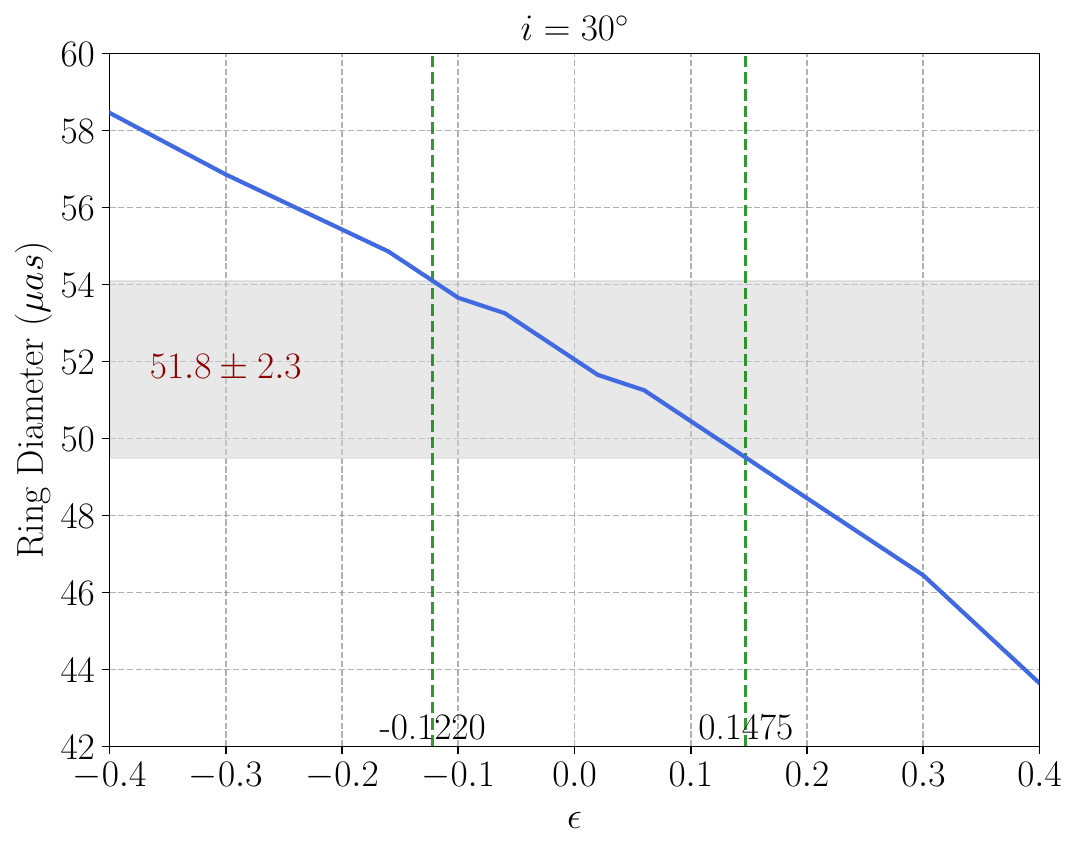}
    \includegraphics[width=0.3\textwidth]{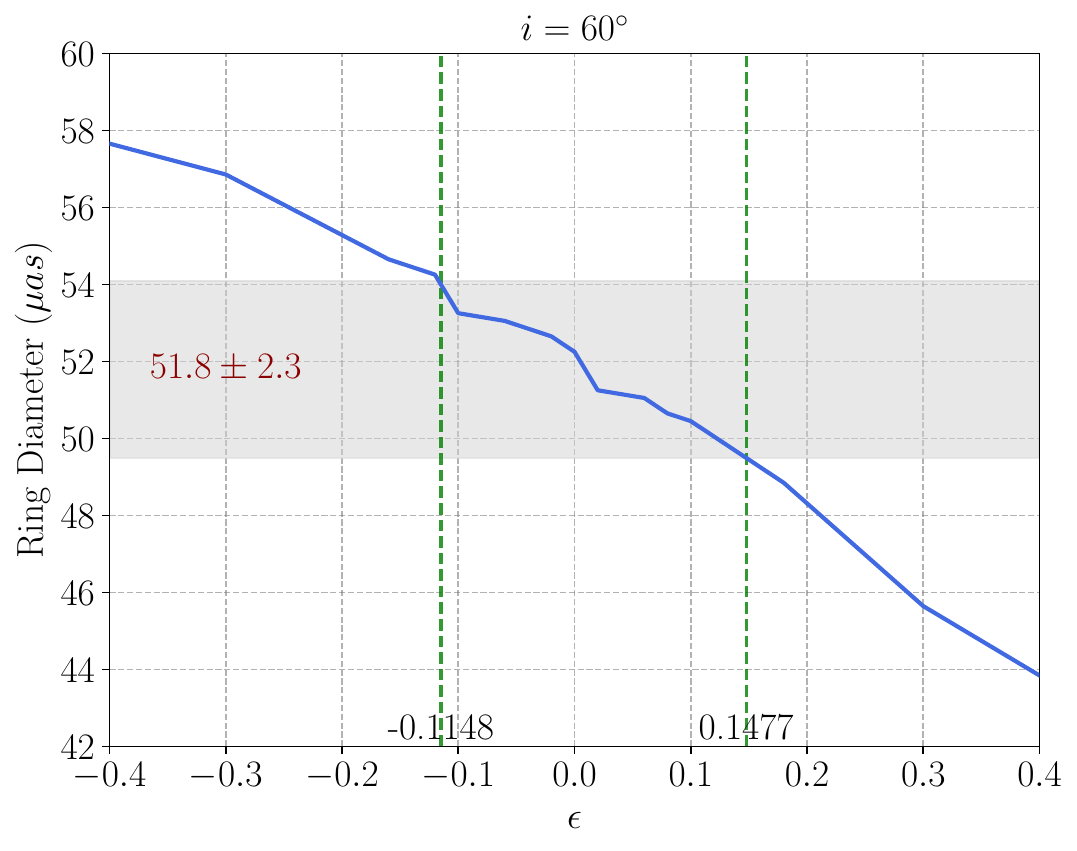}
\caption{This figure illustrates the relationship between the black hole photon ring diameter (in microarcseconds, $\rm \mu as$) and the parameter $\epsilon$  under an observation inclination of $5^\circ$, $30^\circ$ and $60^\circ$. The blue solid line represents the theoretically calculated photon ring diameter, while the light gray shaded area indicates the observational value  $51.8\pm2.3 \rm \mu as$ with its error range. The parameter $\epsilon$ is constraint to be  $-0.1129 \le \epsilon \le 0.1582$ for an observation inclination of $5^\circ$, $-0.1220 \le \epsilon \le 0.1475$ for $30^\circ$, and $-0.1148 \le \epsilon \le 0.1477$ for $60^\circ$.}
    \label{rd}
\end{figure*}

We further extract the photon ring diameter from the flux profiles and present its dependence on $\epsilon$ in Fig.~\ref{rd}. By interpolating these results and comparing them with the measured angular diameter $51.8 \pm 2.3\,\mu\text{as}$ reported by the EHT Collaboration~\cite{EventHorizonTelescope:2022wkp}, we derive observational constraints on the parameter $\epsilon$. Specifically, we obtain:
\begin{itemize}
    \item $-0.1129 \le \epsilon \le 0.1582$ for an inclination angle of $5^\circ$,
    \item $-0.1220 \le \epsilon \le 0.1475$ for $30^\circ$,
    \item $-0.1148 \le \epsilon \le 0.1477$ for $60^\circ$.
\end{itemize}

These bounds are not only consistent with our earlier constraints obtained from the orbital dynamics of the S2 star~\cite{Yan:2024bim}, but also significantly tighter, thanks to the high precision of the EHT observations. The similarity of the constraints across different inclination angles suggests that the photon ring diameter is predominantly governed by the deformation parameter $\epsilon$, while the inclination angle plays only a secondary role in determining its precise value.

\subsection{Polarization in Buchdahl-Inspired Spacetime}

The polarization properties of black holes serve as an important observational probe for investigating the magnetized structure of the accretion flow. Such properties provide insights into the magnetohydrodynamic (MHD) characteristics of the disk-jet system and the influence of strong gravitational fields on polarized light.

The Buchdahl-inspired spacetime introduces a deformation parameter $\epsilon$ to describe an asymptotically flat $\mathcal{R}^2$ spacetime. Studying the effects of this spacetime on polarization modes helps in understanding the coupling mechanisms between strong gravity and accretion disk Magnetohydrodynamics(MHD) processes, and further serves as a test for general relativity in extreme gravity environments. The parameter $\epsilon$, by modifying the underlying spacetime structure, inevitably influences the geodesic motion of photons. However, since the polarization pattern is primarily governed by MHD processes, the influence of $\epsilon$ on polarization is expected to be relatively minor. Nonetheless, variations in $\epsilon$ may lead to subtle changes in polarization signatures, reflecting the intricate interplay between spacetime geometry and disk dynamics. 

We adopt the Stokes parameter formalism defined by the International Astronomical Union (IAU), where $I$ denotes the total intensity representing the radiation energy density, $Q$ and $U$ describe the magnitude and orientation of linear polarization, and $V$ represents the circular polarization component, with its sign related to the direction of Faraday rotation.

Following the Event Horizon Telescope (EHT) Collaboration's standard procedure~\cite{EventHorizonTelescope:2021bee}, we define the following observables:
\bq
|m| = \frac{\sum_i \sqrt{Q_i^2 + U_i^2}}{\sum_i I_i}, 
\eq
\bq
|v| = \frac{\sum_i \left| \frac{V_i}{I_i} \right| I_i}{\sum_i I_i}, 
\eq
where the index $i$ sums over all image pixels, and the results depend on the image resolution. The quantity $|m|$ denotes the polarization fraction, reflecting the degree of magnetic field ordering, while $|v|$ denotes the circular polarization fraction, useful for studying magnetic field distortions and Faraday effects in the accretion disk.

In the strong gravity regime, photon trajectories are bent and affected by the combined influences of plasma-induced Faraday rotation, turbulent magnetic fields in the accretion flow, and frame-dragging due to black hole spin. These effects collectively lead to nontrivial modifications in the polarization structure. Therefore, analyzing how Buchdahl-inspired spacetime affects polarization patterns can reveal the interplay between spacetime geometry and magnetic field dynamics near black holes.

To analyze the symmetry of the polarization structure, we perform a Fourier decomposition of the complex linear polarization field $P = Q + iU$:
\bq
\beta_m = \frac{1}{I_{\text{total}}} \int_0^\infty \int_0^{2\pi} P(\rho, \phi) e^{-im\phi} \, \rho \, d\phi \, d\rho, 
\eq
where $(\rho, \phi)$ are the polar coordinates on the image plane, and $I_{\text{total}}$ is the total intensity flux.

The $\beta_1$ mode characterizes the dipolar asymmetry of the polarization field, associated with angular momentum transport between the disk and jet, and is useful for probing jet magnetic topology. The $\beta_2$ mode is the simplest rotationally symmetric mode, describing second-order symmetry in the polarization distribution. Its phase $\angle \beta_2$ captures the helical pattern of the overall EVPA (Electric Vector Position Angle) structure:
\bq
\text{EVPA} \equiv \frac{1}{2} \arctan\left( \frac{U}{Q} \right). 
\eq

In particular, $\angle \beta_2$ serves as a probe of the spiral structure and pitch angle of the EVPA map. A positive $\angle \beta_2$ indicates a clockwise spiral distribution of the polarization vectors, while a negative $\angle \beta_2$ corresponds to a counterclockwise spiral. Specifically, $\angle \beta_2 = 0^\circ$ represents a radial EVPA pattern, whereas $\angle \beta_2 = 180^\circ$ corresponds to a tangential pattern.

\begin{figure*}[htp]
    \centering
    \subfloat{\includegraphics[width=0.33\textwidth]{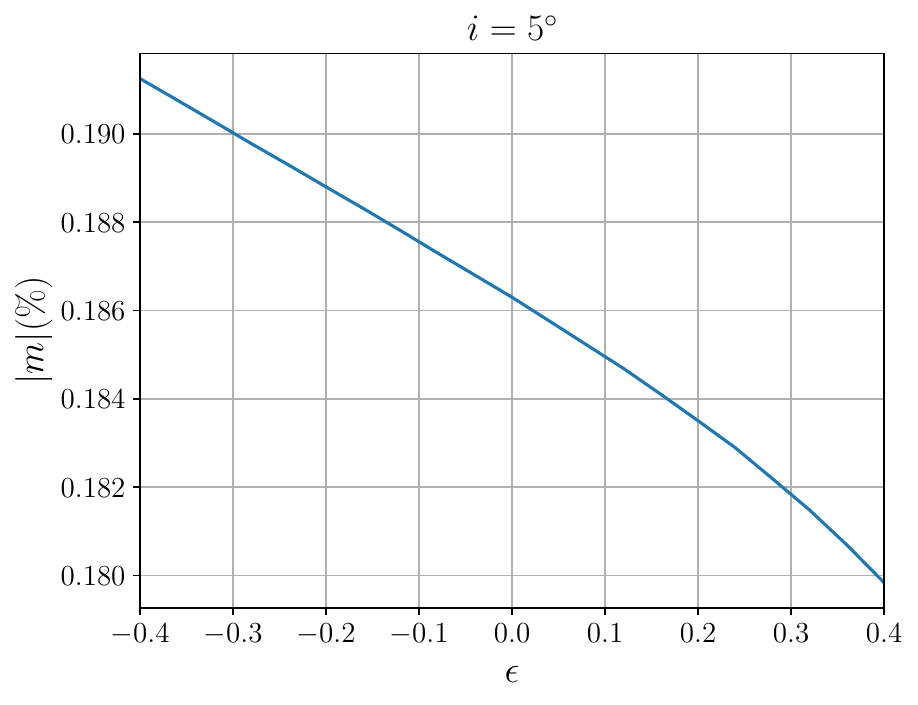}} 
    \subfloat{\includegraphics[width=0.33\textwidth]{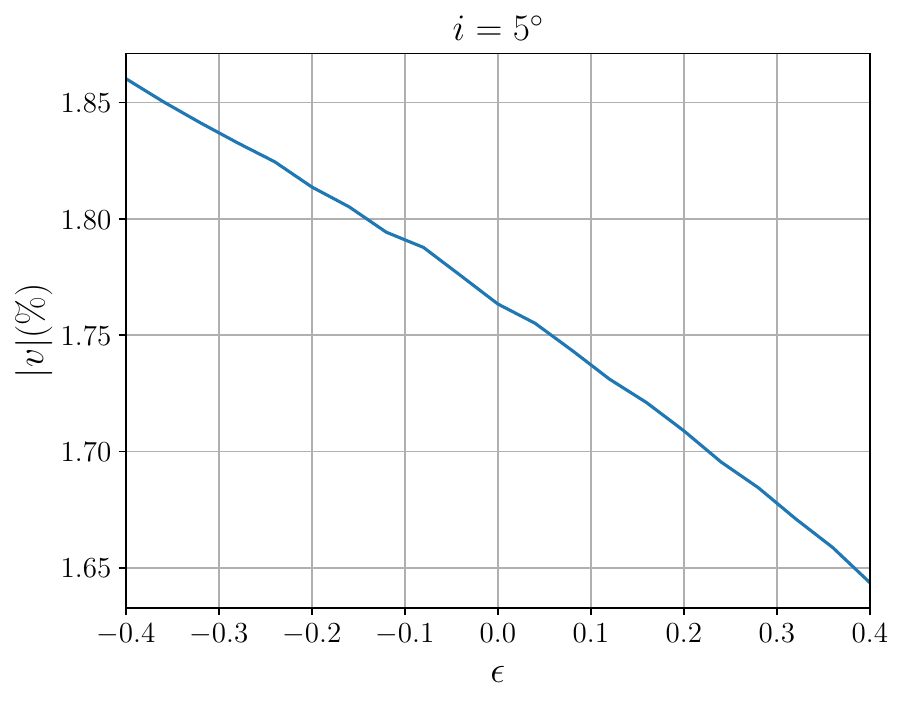}} 
    \subfloat{\includegraphics[width=0.33\textwidth]{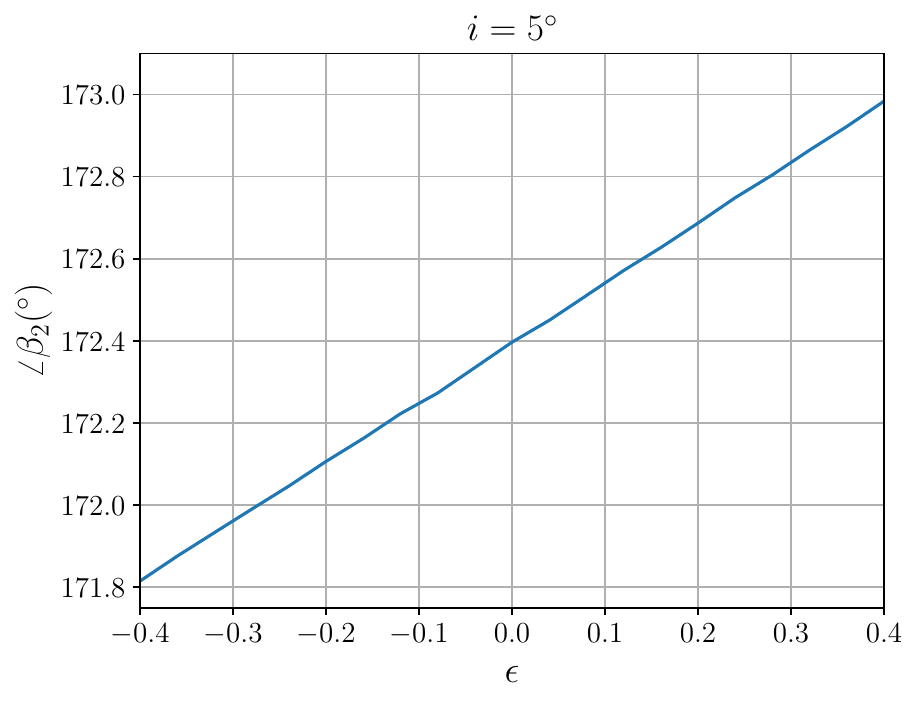}}\\
    \subfloat{\includegraphics[width=0.33\textwidth]{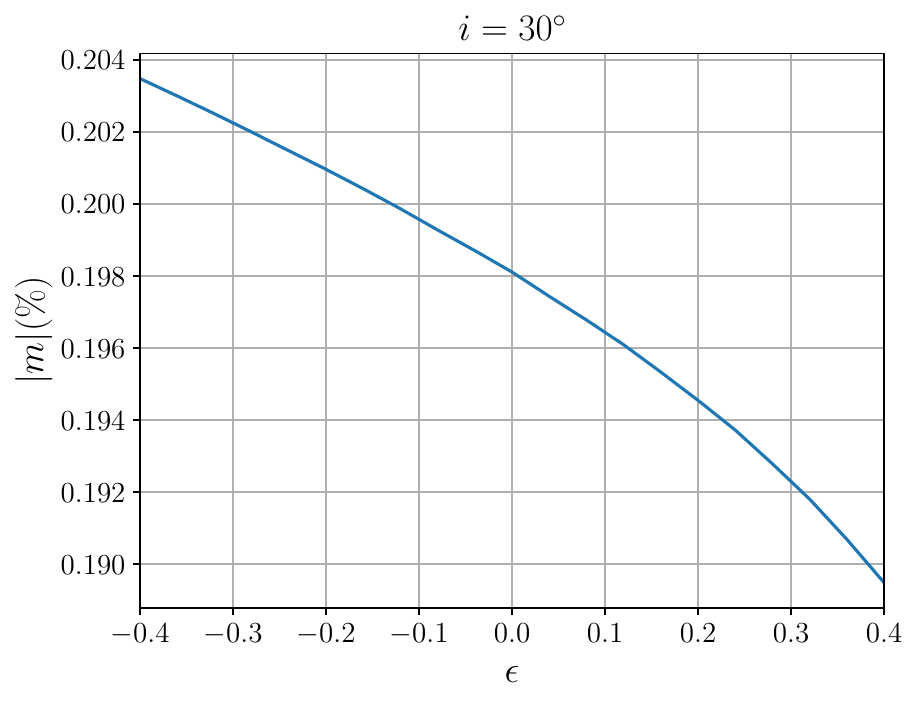}} 
    \subfloat{\includegraphics[width=0.33\textwidth]{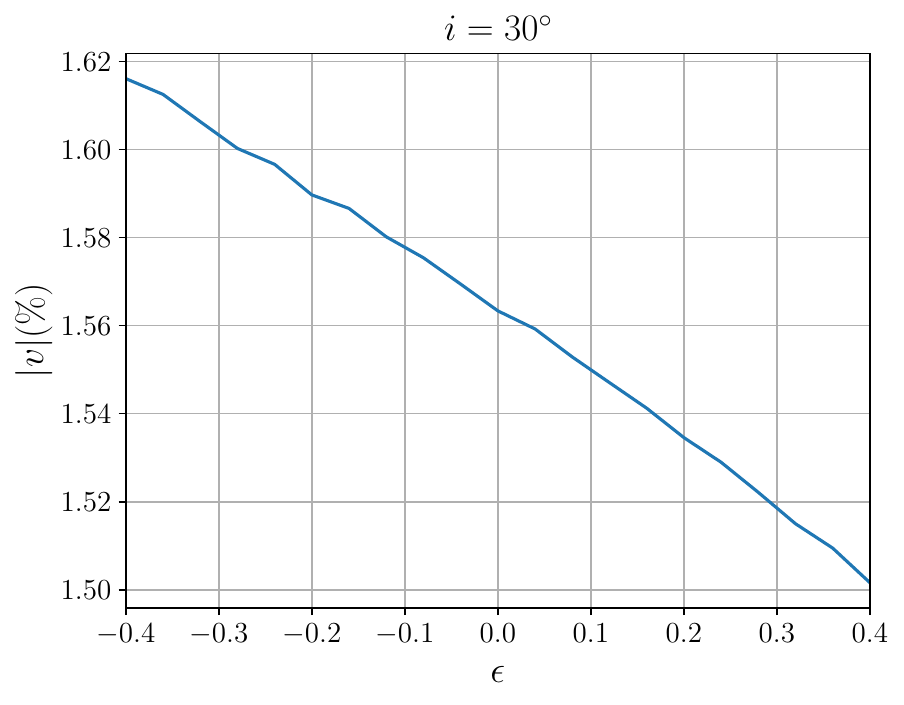}} 
    \subfloat{\includegraphics[width=0.33\textwidth]{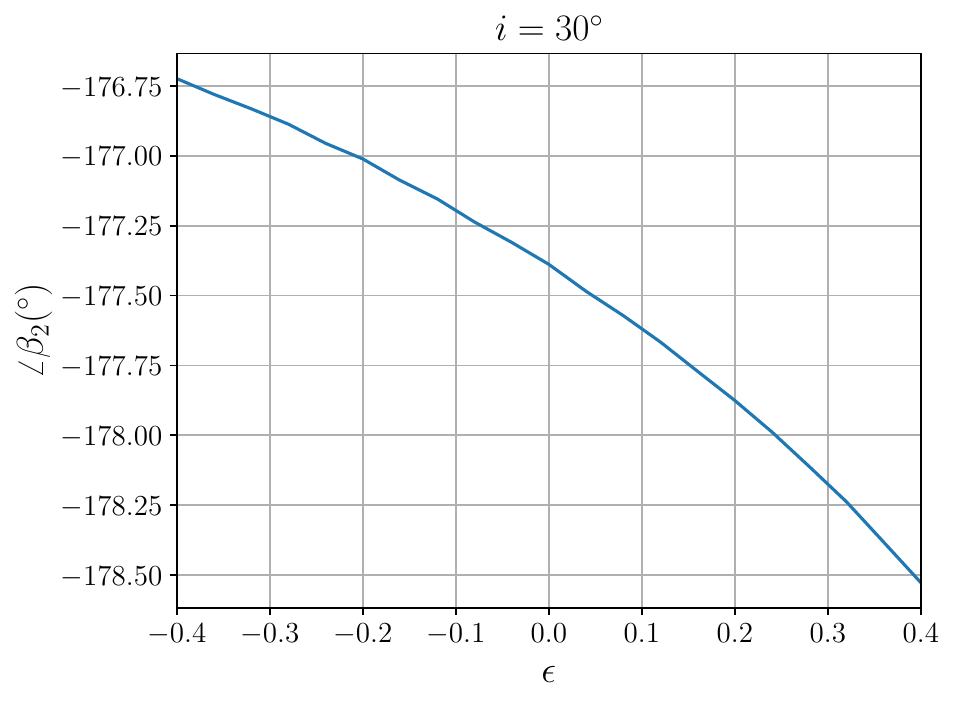}}\\
    \subfloat{\includegraphics[width=0.33\textwidth]{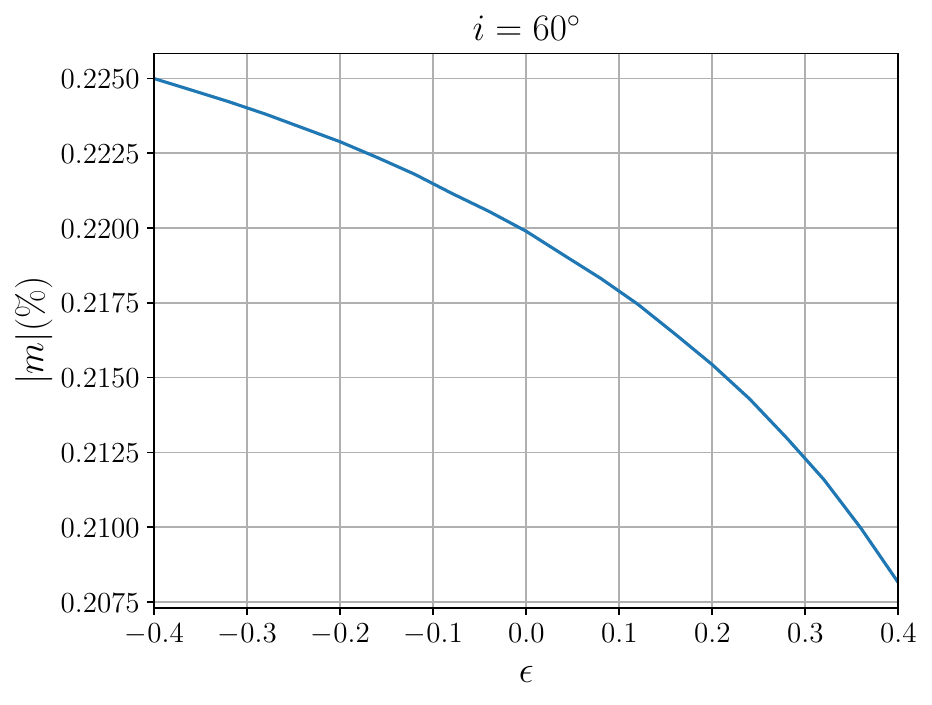}} 
    \subfloat{\includegraphics[width=0.33\textwidth]{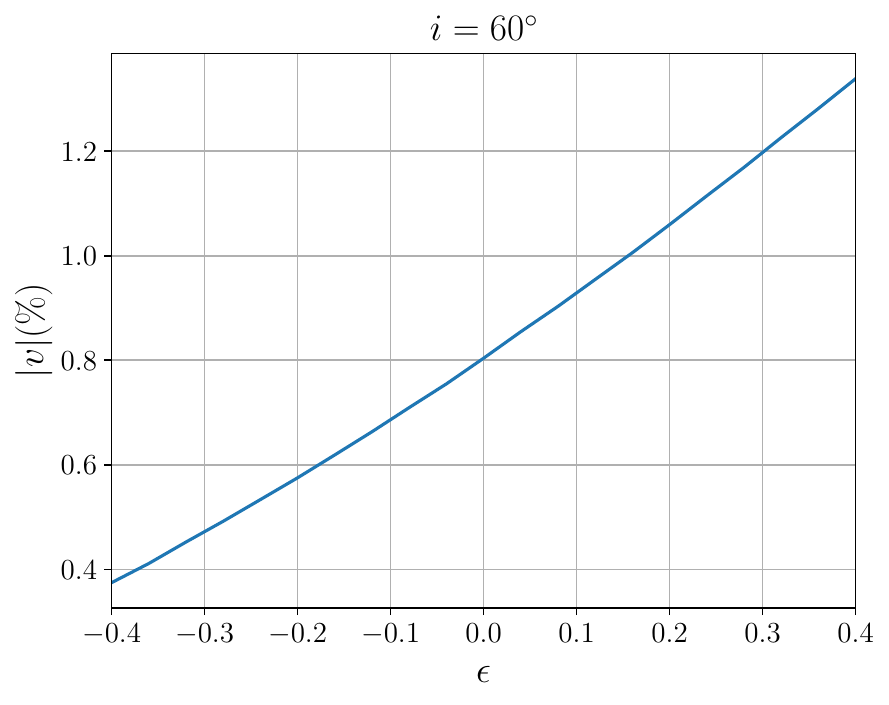}} 
    \subfloat{\includegraphics[width=0.33\textwidth]{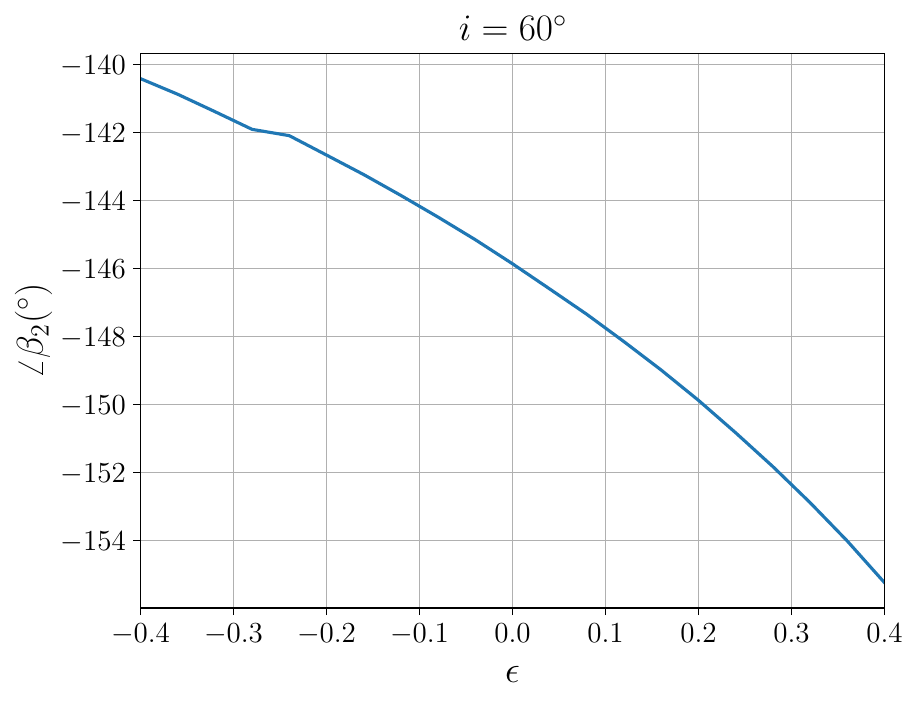}} 
    \caption{This figure show the polarization observables of black hole images in Buchdahl-inspired spacetime for different values of $\epsilon$ under three observation inclination angles: $5^\circ$ (top row), $30^\circ$ (middle row), and $60^\circ$ (bottom row). Each row shows the variation of the polarization fraction $|m|$, circular polarization degree $|v|$, and the EVPA spiral phase angle $\angle \beta_2$ with respect to $\epsilon$. As  $\epsilon$ increases, $|m|$ and $|v|$ generally decrease for all inclination angles, while $\angle \beta_2$ increases. The direction of the EVPA spiral pattern changes from clockwise ($\angle \beta_2 > 0$) at $5^\circ$ to counterclockwise ($\angle \beta_2 < 0$) at higher inclinations. The overall influence of the spacetime deformation parameter $\epsilon$ on polarization is found to be minimal, while the inclination angle has a more noticeable impact on the polarization pattern.}
    \label{polar}
\end{figure*}

As in the previous section, we consider the polarization patterns of black hole images under three different inclination angles: $5^\circ$, $30^\circ$, and $60^\circ$. As shown in the upper row of Fig. \ref{polar}, for an observation inclination angle of $5^\circ$, both $|m|$ and $|v|$ decrease with increasing $\epsilon$, while $\angle \beta_2$ increases with increasing $\epsilon$. Since $\angle \beta_2 > 0$, the polarization electric vectors exhibit a clockwise spiral pattern. Although $|m|$, $|v|$, and $\angle \beta_2$ show varying degrees of change with respect to the parameter $\epsilon$, the overall variation is relatively small, indicating that the gravitational system has a negligible effect on the polarization of the black hole shadow.

Subsequently, we computed the trends of $|m|$, $|v|$, and $\angle \beta_2$ with respect to $\epsilon$ at an observation inclination angle of $30^\circ$, as shown in the middle row of Figs. \ref{polar}. The results show that $|m|$ and $|v|$ continue to decrease with increasing $\epsilon$, while $\angle \beta_2$ continues to increase. Similar to the $5^\circ$ case, the magnitude of these variations remains very small, further indicating that the effect of the gravitational system on polarization is minimal. However, in this case $\angle \beta_2 < 0$, indicating that the polarization electric vectors follow a counterclockwise spiral distribution.

Finally, we analyzed the case of an observation inclination angle of $60^\circ$, as shown in the bottom row of  Figs. \ref{polar}. In this scenario, $|m|$ decreases with increasing $\epsilon$, whereas both $|v|$ and $\angle \beta_2$ increase with increasing $\epsilon$. Since $\angle \beta_2 < 0$, the polarization electric vectors in this case also exhibit a counterclockwise spiral distribution.

By comparing the polarization distributions of black hole images for inclination angles of $5^\circ$, $30^\circ$, and $60^\circ$, we find that the parameter $\epsilon$ has a minimal effect on the polarization structure. In combination with our analysis of the photon ring diameter in the previous subsection, we conclude that the perturbations to EVPA caused by spacetime deformation mainly arise from the gravitational field's influence on photon trajectories. The dominant factor determining the EVPA distribution remains the intrinsic magnetization structure of the accretion flow. In contrast, the observation inclination angle significantly affects the observed polarization distribution around the black hole.

In black hole imaging, the polarization structure reveals critical information about the magnetic field configuration and plasma dynamics near the event horizon. The observer's inclination also affect the polarization pattern of the image. As the observer's inclination angle increases, the polarization image undergoes significant changes due to the combined effects of projection, relativistic beaming, and gravitational lensing.

First, with increasing inclination, the projection of the magnetic field onto the observer's image plane becomes distorted, leading to noticeable variations in the polarization distribution. Second, relativistic motion enhances the emission from the side of the accretion flow approaching the observer, further amplifying asymmetries in the polarization map.

At the same time, as light propagates through the curved spacetime near the black hole, the polarization vectors undergo parallel transport, which can result in Faraday rotation during propagation. The extent of this rotation is closely related to the light path, which is itself influenced by the observer's viewing angle. Different inclinations lead to varying path lengths for light rays reaching different parts of the image, thereby affecting the final polarization distribution observed on the screen.

\section{CONCLUSION}

In this paper, we conduct an in-depth study of the $R^2$ gravity model based on the Buchdahl-inspired metric, and investigate the impact of the deformation parameter $\epsilon$ on black hole images using observational data of Sgr A* from the Event Horizon Telescope. 

First, the core algorithm for black hole imaging is based on the ipole simulation, which employs a semi-analytic method to directly solve the photon geodesic equations and describes the polarization radiation field of photons using the Stokes parameters. We introduce the parameter $\epsilon$ from the Buchdahl-inspired metric to quantify spacetime deformation and use this model to simulate changes in the black hole shadow. The study reveals that as $\epsilon$ increases, the black hole shadow and the angular diameter of the photon ring decrease, which is consistent with our expectations.

By comparing the simulated photon ring diameter with the EHT observational results, we preliminarily constrain the range of $\epsilon$ and further validate the effectiveness of the $R^2$ gravity model. 
Under different observation inclination, $\epsilon$ is constraint to be $-0.1129 \le \epsilon \le 0.1582$ for an inclination angle of $5^\circ$, $-0.1220 \le \epsilon \le 0.1475$ for $30^\circ$ and $-0.1148 \le \epsilon \le 0.1477$ for $60^\circ$. These results are consistent with the constraint $-0.6690 \le \epsilon \le 0.4452$ obtained from previous observations of the S2 star but provides a more precise limit. By comparing the photon ring diameters under different observation inclination angles, we find that different inclination angles have a minimal effect on the observed photon ring diameter.

Furthermore, the sensitivity analysis of polarization properties to spacetime deformation shows that although the polarization angle $\angle \beta_2$, linear polarization fraction $|m|$, and circular polarization fraction $|v|$ change slightly as $\epsilon$ increases, the magnitudes of these variations are relatively small, indicating that gravitational effects have a limited impact on the polarization characteristics of the black hole shadow. The polarization changes primarily arise from the magnetohydrodynamic structure of the accretion disk, rather than direct gravitational effects. Additionally, different observation inclination angles significantly influence the observed polarization distribution near the black hole.

In conclusion, the research in this paper demonstrates that the $R^2$ gravity theory can describe black hole spacetime variations to some extent. By comparing with observational results of Sgr A*, especially 
the photon ring diameter, it provides effective constraints on the free parameter $\epsilon$ of the model. This offers an important theoretical basis and experimental framework for the future exploration of more complex gravitational theories.

\section{Acknowledgments}

This work is supported by the National Natural Science Foundation of China under Grants No. 12275238, the Zhejiang Provincial Natural Science Foundation of China under Grants No. LR21A050001 and No. LY20A050002, and the National Key Research and Development Program of China under Grant No. 2020YFC2201503. I am grateful to Hong-Xuan Jiang from Shanghai Jiao Tong University for fruitful discussions and helpful comments.


\begin{thebibliography}{199}

\bibitem{Will:2014kxa}
C.~M.~Will,
The Confrontation between General Relativity and Experiment,
Living Rev. Rel. \textbf{17}, 4 (2014),
doi:10.12942/lrr-2014-4,
[arXiv:1403.7377 [gr-qc]].

\bibitem{Park:2017zgd}
R.~S.~Park, W.~M.~Folkner, A.~S.~Konopliv, J.~G.~Williams, D.~E.~Smith and M.~T.~Zuber,
Precession of Mercury\textquoteright{}s Perihelion from Ranging to the MESSENGER Spacecraft,
Astron. J. \textbf{153}, no.3, 121 (2017),
doi:10.3847/1538-3881/aa5be2.

\bibitem{Fomalont:2009zg}
E.~Fomalont, S.~Kopeikin, G.~Lanyi and J.~Benson,
Progress in Measurements of the Gravitational Bending of Radio Waves Using the VLBA,
Astrophys. J. \textbf{699}, 1395-1402 (2009),
doi:10.1088/0004-637X/699/2/1395,
[arXiv:0904.3992 [astro-ph.CO]].

\bibitem{Cassini}
Bertotti, B., Iess, L., Tortora, P., A test of general relativity using radio links with the Cassini spacecraft, Nature 425, 374-376 (2003),
doi.org/10.1038/nature01997.

\bibitem{Stairs:2003eg}
I.~H.~Stairs,
Testing general relativity with pulsar timing,
Living Rev. Rel. \textbf{6}, 5 (2003),
doi:10.12942/lrr-2003-5,
[arXiv:astro-ph/0307536 [astro-ph]].

\bibitem{EHT1}
K.~Akiyama \textit{et al.} [Event Horizon Telescope], {First M87 Event Horizon Telescope Results. I. The Shadow of the Supermassive Black Hole}, Astrophys. J. Lett. \textbf{875},
L1, (2019) [arXiv:1906.11238 [astro-ph.GA]].

\bibitem{EHT2}
K.~Akiyama \textit{et al.} [Event Horizon Telescope], {First M87 Event Horizon Telescope Results. IV. Imaging
	the Central Supermassive Black Hole}, Astrophys. J. Lett. \textbf{875},
L4, (2019).

\bibitem{EHT3}
K.~Akiyama \textit{et al.} [Event Horizon Telescope], {First M87 Event Horizon Telescope Results. V. Physical
	Origin of the Asymmetric Ring}, Astrophys. J. Lett. \textbf{875},
L5, (2019).

\bibitem{EHT4}
K.~Akiyama \textit{et al.} [Event Horizon Telescope], {First M87 Event Horizon Telescope Results. VI. The Shadow
	and Mass of the Central Black Hole}, Astrophys. J. Lett. \textbf{875},
L6, (2019).

\bibitem{EventHorizonTelescope:2021bee}
K.~Akiyama \textit{et al.} [Event Horizon Telescope],
First M87 Event Horizon Telescope Results. VII. Polarization of the Ring,
Astrophys. J. Lett. \textbf{910}, no.1, L12 (2021),
doi:10.3847/2041-8213/abe71d, [arXiv:2105.01169 [astro-ph.HE]].

\bibitem{EventHorizonTelescope:2022wkp}
K.~Akiyama \textit{et al.} [Event Horizon Telescope],
First Sagittarius A* Event Horizon Telescope Results. I. The Shadow of the Supermassive Black Hole in the Center of the Milky Way,
Astrophys. J. Lett. \textbf{930}, no.2, L12 (2022),
doi:10.3847/2041-8213/ac6674,
[arXiv:2311.08680 [astro-ph.HE]].

\bibitem{gws}
B.~P.~Abbott \textit{et al.} [LIGO Scientific and Virgo],
Observation of Gravitational Waves from a Binary Black Hole Merger,
Phys. Rev. Lett. \textbf{116}, no.6, 061102 (2016)
[arXiv:1602.03837 [gr-qc]].

\bibitem{gws2}
B.~P.~Abbott \textit{et al.} [LIGO Scientific and Virgo],
Properties of the Binary Black Hole Merger GW150914,
Phys. Rev. Lett. \textbf{116}, no.24, 241102 (2016)
[arXiv:1602.03840 [gr-qc]].

\bibitem{Gaume-2016}
L. Alvarez-Gaume, A. Kehagias, C. Kounnas, D. L\"ust, and A. Riotto, 
Aspects of quadratic gravity, 
Fortschr.$\,$Phys. \textbf{64}, 176 (2016).

\bibitem{Buchdahl-1962}
H.$\,$A. Buchdahl, 
On the Gravitational Field Equations Arising from the Square of the Gaussian Curvature,
Nuovo Cimento \textbf{23}, 141 (1962).

\bibitem{Nguyen-2022-Buchdahl}
H.$\,$K. Nguyen, 
Beyond Schwarzschild-de Sitter spacetimes: I. A new exhaustive class of metrics inspired by Buchdahl for pure $\mathcal{R}^{2}$ gravity in a compact form, Phys.$\,$Rev.$\,$D \textbf{106}, 104004 (2022).

\bibitem{Nguyen-2022-Lambda0}
H.$\,$K. Nguyen, 
Beyond Schwarzschild-de Sitter spacetimes: II. An exact non-Schwarzschild metric in pure $\mathcal{R}^{2}$ gravity and new anomalous properties of $\mathcal{R}^{2}$ spacetimes,
Phys.$\,$Rev.$\,$D \textbf{107}, 104008 (2023) [arXiv:2211.03542 [gr-qc]].

\bibitem{Nguyen-2023-essay}
H.$\,$K. Nguyen, 
Buchdahl-inspired spacetimes and wormholes: Unearthing Hans Buchdahl's other `hidden' treasure trove, Int. J. Mod. Phys. D, 2342007 (2023) [arXiv:2305.08163 [gr-qc]].

\bibitem{2023-WH}
H.$\,$K. Nguyen and M. Azreg-A\"inou, 
Traversable Morris-Thorne-Buchdahl wormholes in quadratic gravity, 
Eur.$\,$Phys.$\,$J.$\,$C \textbf{83}, 626 (2023) [arXiv:2305.04321 [gr-qc]].

\bibitem{2023-axisym}
M. Azreg-A\"inou and H.$\,$K. Nguyen, 
A stationary axisymmetric vacuum solution for pure $\mathcal{R}^{2}$ gravity, Phys. Scr. \textbf{98}, 125025 (2023) [arXiv:2304.08456 [gr-qc]].

\bibitem{taohoang}
T.~Zhu, H.~K.~Nguyen, M.~Azreg-A\"\i{}nou and M.~Jamil,
Observational tests of asymptotically flat ${{\mathcal {R}}}^{2}$ spacetimes,
Eur. Phys. J. C \textbf{84}, 330 (2024).

\bibitem{Yan:2024bim}
J.~M.~Yan, T.~Zhu, M.~Azreg-A\"\i{}nou, M.~Jamil and H.~K.~Nguyen,
Observational test of \ensuremath{\mathcal{R}}$^{2}$ spacetimes with the S2 star in the Milky Way galactic center,
JCAP \textbf{07}, 071 (2024),
doi:10.1088/1475-7516/2024/07/071,
[arXiv:2405.10559 [gr-qc]].

\bibitem{Psaltis:2018xkc}
D.~Psaltis,
Testing General Relativity with the Event Horizon Telescope,
Gen. Rel. Grav. \textbf{51}, no.10, 137 (2019),
doi:10.1007/s10714-019-2611-5,
[arXiv:1806.09740 [astro-ph.HE]].

\bibitem{Vagnozzi:2022moj}
S.~Vagnozzi, R.~Roy, Y.~D.~Tsai, L.~Visinelli, M.~Afrin, A.~Allahyari, P.~Bambhaniya, D.~Dey, S.~G.~Ghosh and P.~S.~Joshi, \textit{et al.}
Horizon-scale tests of gravity theories and fundamental physics from the Event Horizon Telescope image of Sagittarius A,
Class. Quant. Grav. \textbf{40}, no.16, 165007 (2023),
doi:10.1088/1361-6382/acd97b,
[arXiv:2205.07787 [gr-qc]].

\bibitem{Perlick:2021aok}
V.~Perlick and O.~Y.~Tsupko,
Calculating black hole shadows: Review of analytical studies,
Phys. Rept. \textbf{947}, 1-39 (2022),
doi:10.1016/j.physrep.2021.10.004,
[arXiv:2105.07101 [gr-qc]].

\bibitem{Allahyari:2019jqz}
A.~Allahyari, M.~Khodadi, S.~Vagnozzi and D.~F.~Mota,
Magnetically charged black holes from non-linear electrodynamics and the Event Horizon Telescope,
JCAP \textbf{02}, 003 (2020),
doi:10.1088/1475-7516/2020/02/003,
[arXiv:1912.08231 [gr-qc]].

\bibitem{Gralla:2020srx}
S.~E.~Gralla, A.~Lupsasca and D.~P.~Marrone,
The shape of the black hole photon ring: A precise test of strong-field general relativity,
Phys. Rev. D \textbf{102}, no.12, 124004 (2020),
doi:10.1103/PhysRevD.102.124004,
[arXiv:2008.03879 [gr-qc]].

\bibitem{Cardoso:2018ptl}
V.~Cardoso, M.~Kimura, A.~Maselli and L.~Senatore,
Black Holes in an Effective Field Theory Extension of General Relativity,
Phys. Rev. Lett. \textbf{121}, no.25, 251105 (2018),
[erratum: Phys. Rev. Lett. \textbf{131}, no.10, 109903 (2023)],
doi:10.1103/PhysRevLett.121.251105,
[arXiv:1808.08962 [gr-qc]].

\bibitem{Volkel:2020xlc}
S.~H.~V\"olkel, E.~Barausse, N.~Franchini and A.~E.~Broderick,
EHT tests of the strong-field regime of general relativity,
Class. Quant. Grav. \textbf{38}, no.21, 21LT01 (2021),
doi:10.1088/1361-6382/ac27ed,
[arXiv:2011.06812 [gr-qc]].

\bibitem{Yan:2019hxx}
S.~F.~Yan, C.~Li, L.~Xue, X.~Ren, Y.~F.~Cai, D.~A.~Easson, Y.~F.~Yuan and H.~Zhao,
Testing the equivalence principle via the shadow of black holes,
Phys. Rev. Res. \textbf{2}, no.2, 023164 (2020),
doi:10.1103/PhysRevResearch.2.023164,
[arXiv:1912.12629 [astro-ph.CO]].

\bibitem{Chen:2021lvo}
Y.~Chen, Y.~Liu, R.~S.~Lu, Y.~Mizuno, J.~Shu, X.~Xue, Q.~Yuan and Y.~Zhao,
Stringent axion constraints with Event Horizon Telescope polarimetric measurements of M87*,
Nature Astron. \textbf{6}, no.5, 592-598 (2022),
doi:10.1038/s41550-022-01620-3,
[arXiv:2105.04572 [hep-ph]].

\bibitem{Chen:2022nbb}
Y.~Chen, R.~Roy, S.~Vagnozzi and L.~Visinelli,
Superradiant evolution of the shadow and photon ring of Sgr A\ensuremath{\star},
Phys. Rev. D \textbf{106}, no.4, 043021 (2022),
doi:10.1103/PhysRevD.106.043021,
[arXiv:2205.06238 [astro-ph.HE]].


\bibitem{Afrin:2022ztr}
M.~Afrin, S.~Vagnozzi and S.~G.~Ghosh,
Tests of Loop Quantum Gravity from the Event Horizon Telescope Results of Sgr A*,
Astrophys. J. \textbf{944}, no.2, 149 (2023),
doi:10.3847/1538-4357/acb334,
[arXiv:2209.12584 [gr-qc]].

\bibitem{Kuang:2022ojj}
X.~M.~Kuang, Z.~Y.~Tang, B.~Wang and A.~Wang,
Constraining a modified gravity theory in strong gravitational lensing and black hole shadow observations,
Phys. Rev. D \textbf{106}, no.6, 064012 (2022),
doi:10.1103/PhysRevD.106.064012,
[arXiv:2206.05878 [gr-qc]].

\bibitem{Jiang:2023img}
H.~X.~Jiang, C.~Liu, I.~K.~Dihingia, Y.~Mizuno, H.~Xu, T.~Zhu and Q.~Wu,
Shadows of loop quantum black holes: semi-analytical simulations of loop quantum gravity effects on Sagittarius~A* and M87*,
JCAP \textbf{01}, 059 (2024),
doi:10.1088/1475-7516/2024/01/059,
[arXiv:2312.04288 [gr-qc]].

\bibitem{Hou:2022eev}
Y.~Hou, Z.~Zhang, H.~Yan, M.~Guo and B.~Chen,
Image of a Kerr-Melvin black hole with a thin accretion disk,
Phys. Rev. D \textbf{106}, no.6, 064058 (2022),
doi:10.1103/PhysRevD.106.064058,
[arXiv:2206.13744 [gr-qc]].

\bibitem{Jusufi:2020cpn}
K.~Jusufi, M.~Jamil and T.~Zhu,
Shadows of Sgr A$^{*}$ black hole surrounded by superfluid dark matter halo,
Eur. Phys. J. C \textbf{80}, no.5, 354 (2020),
doi:10.1140/epjc/s10052-020-7899-5,
[arXiv:2005.05299 [gr-qc]].

\bibitem{Liu:2020ola}
C.~Liu, T.~Zhu, Q.~Wu, K.~Jusufi, M.~Jamil, M.~Azreg-A\"\i{}nou and A.~Wang,
Shadow and quasinormal modes of a rotating loop quantum black hole,
Phys. Rev. D \textbf{101}, no.8, 084001 (2020),
[erratum: Phys. Rev. D \textbf{103}, no.8, 089902 (2021)],
doi:10.1103/PhysRevD.101.084001,
[arXiv:2003.00477 [gr-qc]].

\bibitem{Zhu:2019ura}
T.~Zhu, Q.~Wu, M.~Jamil and K.~Jusufi,
Shadows and deflection angle of charged and slowly rotating black holes in Einstein-\AE{}ther theory,
Phys. Rev. D \textbf{100}, no.4, 044055 (2019),
doi:10.1103/PhysRevD.100.044055,
[arXiv:1906.05673 [gr-qc]].

\bibitem{Shi:2024bpm}
H.~Y.~Shi and T.~Zhu,
Polarized image of a synchrotron emitting ring around a static hairy black hole in Horndeski theory,
Eur. Phys. J. C \textbf{84}, no.8, 814 (2024),
doi:10.1140/epjc/s10052-024-13198-3.



\bibitem{MorrisThorne}
M. S. Morris and K. S. Thorne, Wormholes in spacetime
and their for interstellar travel: A tool for teaching general relativity, Am. J. Phys. \textbf{56}, 5 (1988).

\bibitem{CampanelliLousto-1993}M. Campanelli and C. Lousto, Are Black Holes in Brans--Dicke Theory precisely the same as in General Relativity?, Int. J. Mod. Phys. D \textbf{2}, 451 (1993).

\bibitem{Pu:2016qak}
H.~Y.~Pu, K.~Akiyama and K.~Asada,
The Effects of Accretion Flow Dynamics on the Black Hole Shadow of Sagittarius A$^{*}$,
Astrophys. J. \textbf{831}, no.1, 4 (2016),
doi:10.3847/0004-637X/831/1/4,
[arXiv:1608.03035 [astro-ph.HE]].

\bibitem{Moscibrodzka:2017lcu}
M.~Moscibrodzka and C.~F.~Gammie,
ipole \textendash{} semi-analytic scheme for relativistic polarized radiative transport,
Mon. Not. Roy. Astron. Soc. \textbf{475}, no.1, 43-54 (2018),
doi:10.1093/mnras/stx3162,
[arXiv:1712.03057 [astro-ph.HE]].

\end{thebibliography}
\end{document}